# Optimal Signal Selection for Sensors
# (October 2020)

Abdulaziz M. Alqarni, *Member, IEEE,* and Thomas G. Robertazzi, *Fellow, IEEE*

*Abstract*—The focus of this research is sensor applications including radar and sonar. Optimal sensing means achieving the best sensing quality with the least time and energy cost, which allows processing more data. This paper presents novel work by using an integer linear programming "algorithm" to achieve optimal sensing by selecting the best possible number of signals of a type or a combination of multiple types of signals to ensure the best sensing quality considering all given constraints. A solution based on a heuristic algorithm is implemented to improve the computing time performance. What is novel in this solution is the synthesis of an optimized signal mix using information such as but not limited to signal quality, energy and computing time.

*Index Terms*— Sensors, Heuristic algorithm, Integer linear programming, Signal selection and Signal optimization

## I. Introduction

Wireless sensor networks are useful in diverse application areas. Monitoring applications, military applications, mobile commerce, smart offices and environmental science are examples of these application areas. Monitoring applications include medical health monitoring and structural health monitoring. Surveillance, target tracking, counter sniper, and battlefield monitoring are examples of military applications using wireless sensor networks. This developing technology reduces time and effort used in collecting data and monitoring events. It has advantages as well as challenges and shortcomings. For example, it is convenient, flexible, and accurate. On the other hand, robustness, scalability, and security are examples of the challenges for this technology [1,2]. Sensing means the act of collecting information about an object. It may be split into passive sensors that gather radiation that is emitted or reflected by the object. Passive sensors mostly use reflected sunlight as the source of measured radiation. Examples of passive sensors include film photography, infrared and radiometers. Active sensing is when the sensor emits a signal and detects its reflection by the object. RADAR, SONAR and LiDAR are examples of active sensing [3,4,5].

A sensor's sensitivity indicates how much the input quantity affects the sensor's output. For instance, the temperature changes by 1 °C if the mercury in a thermometer moves 1 cm. The sensitivity in that example is 1 cm/°C. Optimal sensing means the best sensing quality with the least time and energy cost, which allow processing more data [4]. Researchers have optimized signal waveforms using various criteria [6,7,8].

In this paper, an integer linear programming "algorithm" is used to achieve optimal sensing by selecting the best possible number of signals of a type or a combination of multiple types of signals to ensure the best sensing quality possible considering all given constraints. A mathematical optimization problem in which the variables are restricted to be integers is called integer programming. In integer linear programming (ILP) the objective function and the constraint decision variables are linear. The main reason for using integer variables when modeling problems as a linear program: The integer variables represent quantities that can only be integer. For example, it is not possible to build 4.7 cars or in this proposal send 2.5 signals. Integer linear programming can be used in many applications areas such as production planning where a possible objective is to maximize the total production, without exceeding the available resources [9,10,11]. Another example is scheduling such as vehicle scheduling in transportation networks. Also, an example is telecommunications networks where the goal of these problems is to design a network of lines to install so that a predefined set of communication requirements are met, and the total cost of the network is minimal. Finally, cellular networks is another application area such as the task of frequency planning in 5G mobile networks. This involves distributing available frequencies across the antennas so that users can be served, and interference is minimized between the antennas [12,13]. In this paper, the solution to this problem is found using problem-based linear programming, solver-based linear programming and a heuristic algorithm.

Section II discusses potential applications of this concept. The mathematical programming formulation of this problem appears in section III. Various examples of the use of the integer linear programming solution approach are given in section IV. The difference between problem based and solver based linear programming is discussed in section V. The heuristic algorithm is presented VI. A comparison of the three approaches is presented in section VII (Testing Methodology) and section VIII (Results). The conclusion is in section IX.

## II. Applications

Two broad potential application areas for this research are listed in this section.

### A. Sensing

This includes radar, sonar, lidar and other remote sensing technologies. It may be desired that an active radar, active sonar or active lidar or other remote sensing technology emits a mix of signals in an instance (which may be periodically repeated) that maximizes the quality of the return signal subject to constraints in developing the signal mix such as, but not limited to, energy and computing time.

Alternately if the signal quality, energy and computing metrics or other metrics used vary with time, the mix of signals may be modified over time to match current conditions (i.e. current values of the signal, energy, computing time or other metrics being considered). Over time the types of signals that can be included in a signal mix may vary and at each instance the signal mix can be optimized using the proposed solution.

This technique also applies to passive sensing systems. Monitored signals may be classified into different signal types and it may be desired to develop an optimal mix of such signal types for purposes such as but not limited to further processing.

### B. Telecommunications

The proposed technique may be applied to telecommunication systems such as but not limited to 4G, 5G and proposed 6G and 7G cellular systems. This involves signals being transmitted/received by base stations or relays to/from users. Technologies used include spatial diversity (the use of multiple antennas which may be in arrays), spatial multiplexing (the use of multiple paths in MIMO (multi input multi output) configurations with a large number of antennas each individually controlled and potentially with embedded radio transceiver components) and beam forming (by phase adjustment in large antenna arrays the signal directivity can be controlled).

All these technologies can make use of optimized signal mixes. In an active mode, the signal quality for individual signals can be predetermined or determined by receivers which inform transmitters of the quality of the signal information they receive. This is a commonly used technique historically for cellular communications. The signal mix optimization can also be done in a passive mode as described above for sensing applications. Also, time varying environments can be handled as described above for sensing.

### III. PROBLEM FORMULATION AND SOLUTION

The goal is to select the mix of signals that provide the maximum quality "Q" in order to have optimal signature selection as shown in equation (1) [16]. We assumed that there are i types of signals. The question is if a sensor is sending out N signals altogether, how many n signals of each type i summing to N give the best solution? Our problem formulation example starts by creating three types of signals "i". Then, the quality "$q_i$", computation time "$t_i$" and energy "$e_i$" specifications per signal were provided for each type. Then, the overall computation time and energy constraints were set as shown in equations (2) and (3) below. The constraints were set so that the total energy does not exceed the energy constraint "$E_c$" and the total computation time does not exceed the time constraint "$T_c$". The sum of the computation time per signals for each type multiplied by the number of signals from that type "$n_i$" for all three types gives the total computation time "T". Similarly, the sum of the energy per signals for each type multiplied by the number of signals from that type for all three types gives the total energy "E". Different cases were created to test this solution with different sets of constraints, and they will be explained in the following section. Finally, the objective function of the algorithm is created to find the optimal number of signals of each type that would result in the maximum total quality. The maximum total quality can be calculated be multiplying the number of selected signals of each type by the quality per signal for that type then summing the results of all types as shown in equation (1).

$$i = 1,2,3$$
$$n_i\,,\,t_i\,,\,e_i\,,Q\,,\,q_i\,,\,E_c,\,E,\,T\,,\,T_c \text{ are positive numbers}$$
$$\text{Max } Q = \sum_i n_i\,q_i \quad (1)$$
$$\sum_i n_i\,t_i \leq T_c \quad (2)$$
$$\sum_i n_i\,e_i \leq E_c \quad (3)$$

Modifications in the specifications and/or constraints result in different optimal number of signals of each type. Other constraints are possible also the organization of the code controls the prioritization of the constraints. The performance will be evaluated in detail in the following section.

### IV. PERFORMANCE EVALUATION

The performance evaluation portion of this solution was done in three stages. The first stage is studying basic cases of constraints. Four different cases were created for exposition. In every case the quality time and energy specifications per signal were provided for each type. Three types of signals were created for this problem. Then, the overall time and energy constraints were set. Finally, the objective of the algorithm is to find the optimal number of signals of each type that would result in the maximum total quality.

### A. Basic Case

For the first case, the quality per signal for the first, second and third type are two, five and ten (the larger "q" the better) and the computation time per signal are three, two and a half and two microseconds as shown in Figure 1 below. Figure 2 shows the relationship between the energy and quality per signal. The energy per signal values for the three types are one hundred, two hundred and three hundred. Then, two constraints were set for the maximum energy and time cost. The constraints were set so the total energy does not exceed one thousand and the total computation time does not exceed twenty-five microseconds. If the solution should consist of one signal type, the maximum quality would be thirty and the selected signals are three signals of the third type. When a mix of signal types is used for the solution, the result shown in Figure 3 is one signal of the first type and three signals of the third type. The selected signals energy cost is one thousand and the total computation time equals nine microseconds. Finally, the maximum total quality for this case equals thirty-two.

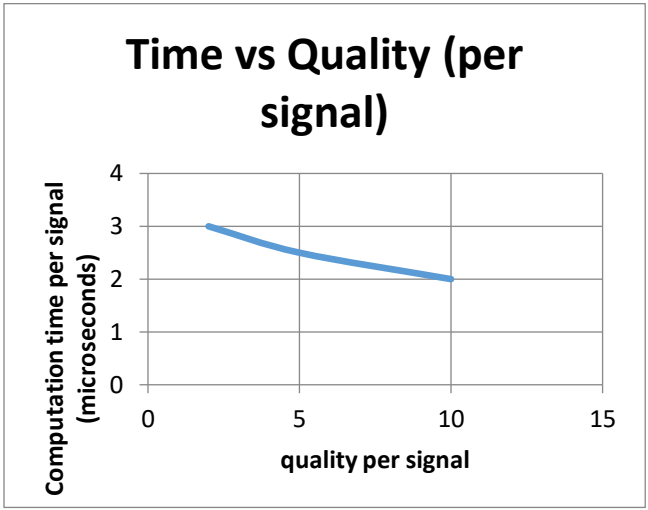

Figure 1: Computational time vs. quality per signal

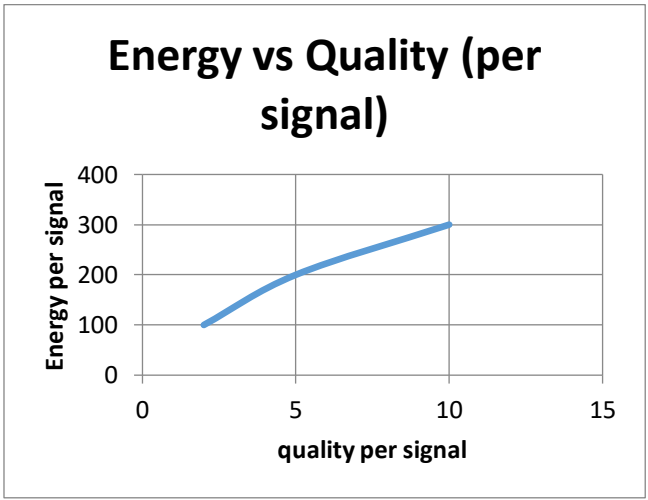

Figure 2: Energy vs. quality per signal

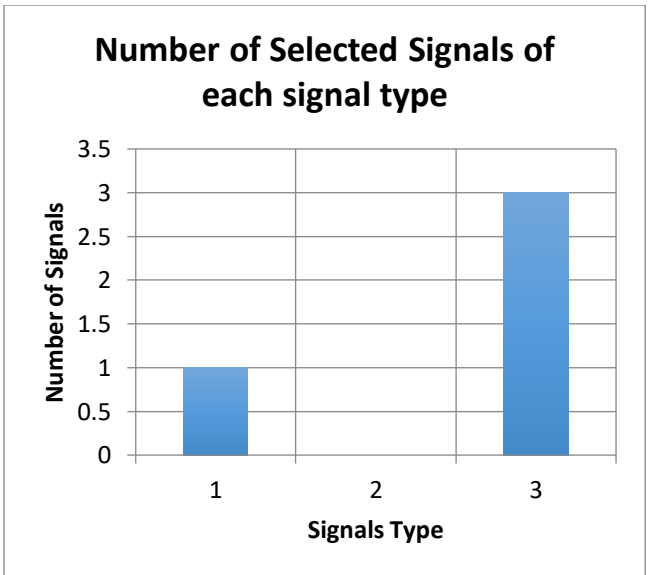

Figure 3: The number of selected signals per signal type

### B. Complex Cases

The second stage is studying more complex cases of linear relationship between energy/computation time and quality per signal. This stage consists of two parts. The first part is combinations of five cases to study the relationship between the energy and quality per signal with constant computation time specifications. Secondly, studying combinations of five cases to find the relationship between the computation time and quality per signal with constant energy specifications.

The first part is a combination of five cases with a linear relationship between energy and quality per signal when the computation time is constant. The quality per signal for the first, second and third signal type are two, five and ten and the computation time per signal are three, two and a half and two microseconds as shown in Table 1 below. Figure 4 shows the linear relationship between the energy and quality per signal type. The energy is increased in each case. Then, two constraints were set for the maximum energy and time cost. The constraints were set so the total energy does not exceed one thousand and the total computation time does not exceed twenty-five microseconds. The selected signals energy cost and the total computation time vary for each case. For the first case, the energy cost is six hundred and the computation time equals twenty-four microseconds. The energy cost for the second case is one thousand and the computation time equals twenty microseconds. The third case energy cost equals one thousand and the computation time is twenty-four microseconds. In the fourth case, nine hundred and ninety is the energy cost and the computation time is eighteen and a half microseconds. Finally, the energy cost for the fifth case equals a thousand and the computation time is six and a half microseconds. The results in Table 2 show that increasing the energy cost leads to a drop in the maximum total quality.

| i | $t_i$ | $q_i$ | E= 5Q | E= 10Q | E= 20Q | E= 30Q | E= 40Q |
|---|---|---|---|---|---|---|---|
| 1 | 3 | 2 | 10 | 20 | 40 | 60 | 80 |
| 2 | 2.5 | 5 | 25 | 50 | 100 | 150 | 200 |
| 3 | 2 | 10 | 50 | 100 | 200 | 300 | 400 |

Table 1: Combination of five cases with linear relationship between energy and quality

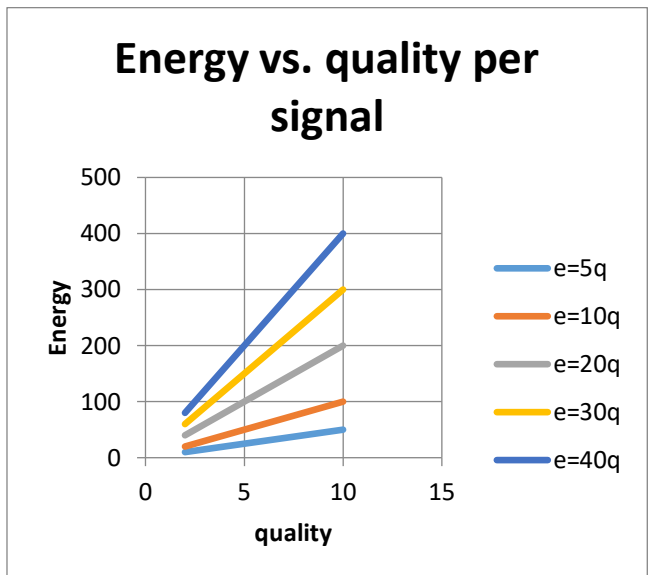

Figure 4: Combination of 5 cases with linear relationship between energy and quality

| Type/case | 1 | 2 | 3 | 4 | 5 |
|---|---|---|---|---|---|
| 1 | 0 | 0 | 5 | 4 | 0 |
| 2 | 0 | 0 | 0 | 1 | 1 |
| 3 | 12 | 10 | 4 | 2 | 2 |
| Quality | 120 | 100 | 50 | 33 | 25 |

Table 2: Number of selected signals per type and the total quality for each case

The second part is a combination of five cases with inverse linear relationship between computation time and quality per signal when the energy is constant. The first case is the inverse linear relationship between computation time and quality when the energy is constant. The quality per signal for the first, second and third signal type are two, five and ten and the energy per signal are one hundred, two hundred and three hundred as shown in Table 3 below. Figure 5 shows the linear relationship between computation time and quality per signal type. The computation time is increased in each case. Then, two constraints were set for the maximum energy and time cost. The constraints were set so the total energy does not exceed one thousand and the total computation time does not exceed twenty-five microseconds. The selected signals energy cost and the total computation time vary for each case. For the first case, the energy cost is one thousand and the computation time equals twenty microseconds. The energy cost for the second case is nine hundred and the computation time equals fifteen microseconds. The third case energy cost equals six hundred and the computation time is twenty microseconds. In the fourth case, three hundred is the energy cost and the computation time is fifteen microseconds. Finally, the energy cost for the fifth case equals one hundred and the computation time is twenty microseconds. The results in Table 4 show that increasing the computation time results in a drop in the maximum total quality.

| $q_i$ | Case 1 t=25/Q | Case 2 t=50/Q | Case 3 t=100/Q | Case 4 t=150/Q | Case 5 t=200/Q |
|---|---|---|---|---|---|
| 2 | 12.5 | 25 | 50 | 75 | 100 |
| 5 | 5 | 10 | 20 | 30 | 40 |
| 10 | 2.5 | 5 | 10 | 15 | 20 |

Table 3: Combination of five cases with inverse linear relationship between Time and Quality

| Type/case | 1 | 2 | 3 | 4 | 5 |
|---|---|---|---|---|---|
| 1 | 1 | 0 | 0 | 0 | 0 |
| 2 | 0 | 0 | 0 | 0 | 0 |
| 3 | 3 | 3 | 2 | 1 | 1 |
| Q | 32 | 30 | 20 | 10 | 10 |

Table 4: number of selected signals per signal type and the total quality for each case

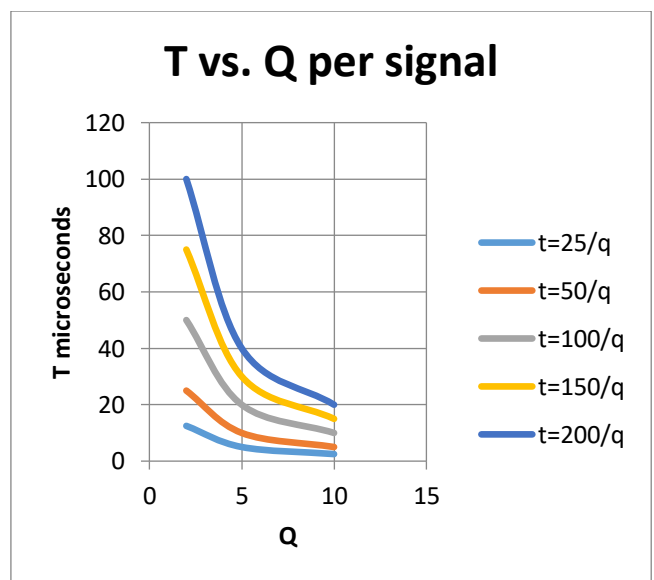

Figure 5: Combination of 5 cases with inverse linear relationship between time and quality per signal

*C. Restriction on Number of Signals*

The third stage is studying the effect of restricting the number of signals per type. In this part the energy and the computation time per signal are constant for each signal type. The quality per signal for the first, second, and third signal type are two, five and ten and the energy per signal are one hundred, two hundred and three hundred. The computation time per signal are five, two and one microseconds. Then, two constraints were set for the maximum energy and time cost. The constraints were set so the total energy does not exceed ten thousand and the total computation time does not exceed one thousand microseconds. Figure 6 shows the number of selected signals per type before restricting the number of signals per type. The total Quality equals three hundred and thirty-two and the selected signals are one signal of the first type and thirty-three signals of the third type.

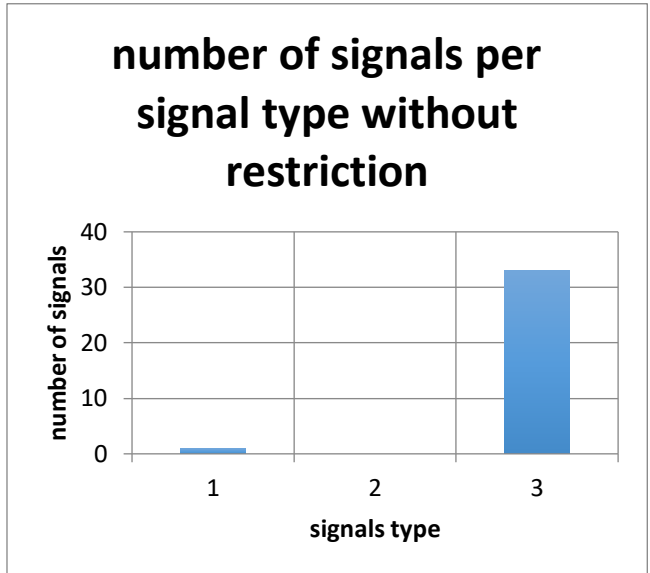

Figure 6: The number of selected signals per signal type without restriction.

The next step is to restrict the number of signals per type as shown in equation (4). Figure 7 shows the number of selected signals per type after adding the constraint "$S_c$", thirty signals per type. The total quality equals three hundred and twenty-five and the selected signals are five signals of the second type and thirty signals of the third type as shown in Figure 7. Then, the number of signals per type constraint is modified to twenty signals per type. For this case, the total quality equals three hundred and the selected signals are twenty signals of the second type and twenty signals of the third type as shown Figure 8. The selected signals per type vary for each case. To verify the changes in results, a third case is done where the number of signals per type constraint is set to ten signals per type. The total quality equals one hundred and seventy and the selected signals are ten signals of each type as shown in Figure 9. In conclusion, restricting the number of signals per signal type affect the maximum total quality.

$$N_i \leq S_c \quad (4)$$

where $S_c$ is a positive integer $0,1,2,3\ldots$

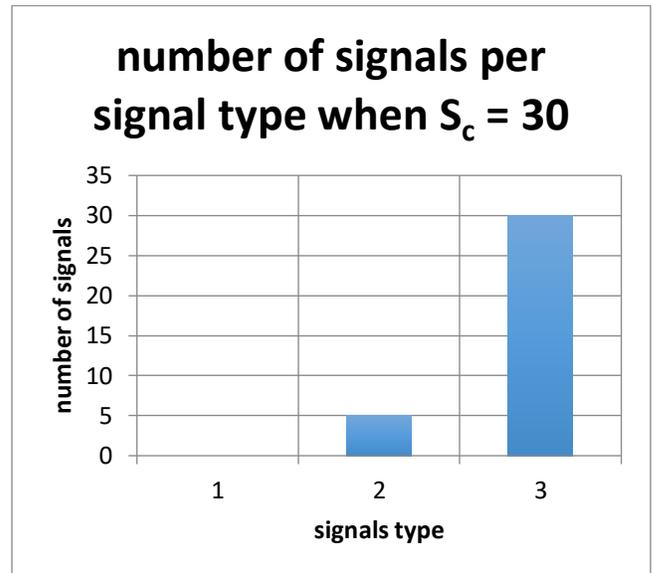

Figure 7: The number of selected signals per signal type when $S_c = 30$.

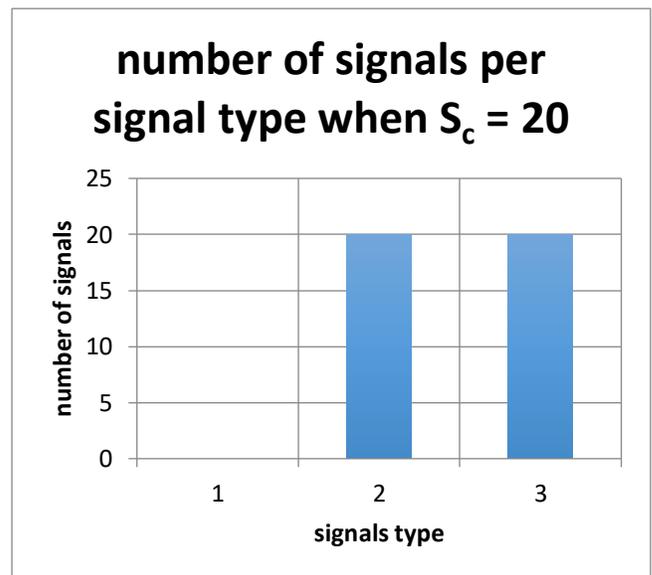

Figure 8: The number of selected signals per signal type when $S_c = 20$.

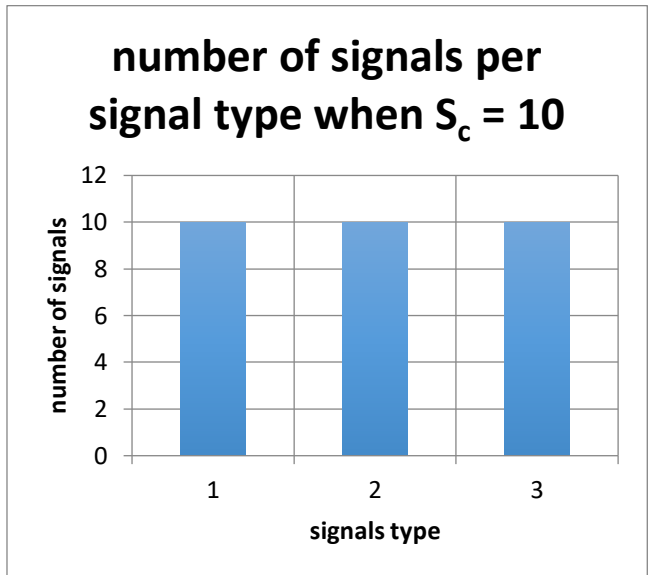

Figure 9: The number of selected signals per signal type when $S_c = 10$.

## V. Comparison between problem based and solver based linear programming

Problem-based and solver-based are two approaches to solving mathematical programming optimization problems in MATLAB. The appropriate approach must be selected before solving a problem. The differences between the two approaches are covered in the following part.

First of all, Problem-Based Optimization Setup is easier to create and debug. The objective and constraints are represented symbolically. It requires translation from problem form to matrix form, resulting in a longer solution time. It does not allow direct inclusion of the gradient or Hessian.

The second approach is solver-based optimization. The problem setup is harder to create and debug. The objective and constraints are represented as functions or matrices. It does not require translation from problem form to matrix form, resulting in a shorter solution time. It allows direct inclusion of gradient or Hessian. Also, it allows use of a Hessian multiply function or Jacobian multiply function to save memory in large problems [14].

The two approaches produce solutions of the same quality. Theoretically, the solver-based solution can improve the performance. This is because the objective and constraints are represented as functions or matrices in solver-based solution. That representation eliminates the translation from problem form to matrix form which allow a shorter solution time.

## VI. Heuristic algorithms solution

A heuristic algorithm can be a faster and more efficient method to solve a problem. Heuristic algorithms are useful to find approximate solutions when it is sufficient and exact solutions are computationally expensive. Heuristic algorithms are commonly employed to solve the Knapsack Problem; a problem related to the problem of this paper. In the knapsack problem, heuristics are used to find the maximum value by grouping a given set of items while being under a certain limit and this solution technique is known as the Greedy Approximation Algorithm. It starts by sorting the items based on their value per unit. Then, it adds the items with the highest value per unit as long as there is still space remaining [15]. Similarly, heuristic algorithm is used to optimize signal selection under a specific set of constraints.

Solving a given problem using a heuristic algorithm could have trade-offs such as optimality, completeness and accuracy. That leads to a question, is the heuristic solution good enough? When multiple solutions exist for a given problem, the following questions can be used to evaluate the solution found by heuristic algorithm; does the heuristic give the best solution? Can the heuristic find all possible solutions? Is this the fastest method for solving this type of problem? A heuristic algorithm was implemented and compared to the previous approaches to answer these questions. The first step is to sorts the signal types. To determine the order of signal types the following expression was used to calculate the quality of a certain signal type

$$\text{Type quality} = q_i/(e_i * t_i) \quad (5)$$

After the signal types are sorted, the algorithm will calculate the best number of signals of each type to find the maximum quality achieved under a specific set of constraints. The same problem formulation in the section II of this paper is used for implementing and testing the heuristic algorithm solution. Finally, the program run time will be used to compare problem-based and solver-based solutions with the heuristic algorithm.

Two different cases were created to study the run time for each program. The first case will be done utilizing only four signal types and the other case will be done utilizing all seven signal types shown in Table 5. Problem-based, solver-based and heuristic algorithm solutions will be used to find the maximum quality utilizing four different signal types. The quality per signal, the energy per signal and the computation time per signal for each signal type are shown in Table 5 below.

For the first case, the quality per signal for each type are five, two, ten and seven. The energy per signal for the utilized four types are two hundred, one hundred, three hundred and two hundred and fifty. Figure 10 shows the relationship between the energy and quality per signal. The computation time per signal are two and five tenths, three, two, and two and three tenths microseconds. The relationship between the computation time and the quality per signal is shown in Figure 11 below.

For the second case, each solution will be done using seven signal types. Three more types of signals were created for this problem. All the parameters for signals types 5, 6 and 7 are provided in Table 5. The quality per signal for each type are twelve, six and one. ten and seven. The energy per signal for the new three types are three hundred and fifty, two hundred and twenty-five and fifty. Figure 10 shows the relationship between the energy and quality per signal. The computation time per signal are one and a half, two and four tenths and three and a half microseconds. The relationship between the computation time and the quality per signal is shown in Figure 11 below.

| i | $q_i$ | $e_i$ | $t_i$ |
|---|---|---|---|
| 1 | 5 | 200 | 2.5 |
| 2 | 2 | 100 | 3 |
| 3 | 10 | 300 | 2 |
| 4 | 7 | 250 | 2.3 |
| 5 | 12 | 350 | 1.5 |
| 6 | 6 | 225 | 2.4 |
| 7 | 1 | 50 | 3.5 |

Table 5: Values of parameters used for testing

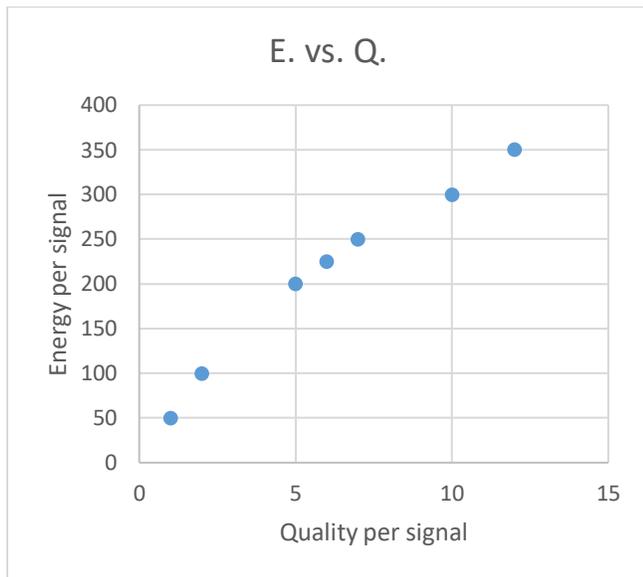

Figure 10: The relationship between the energy and quality per signal

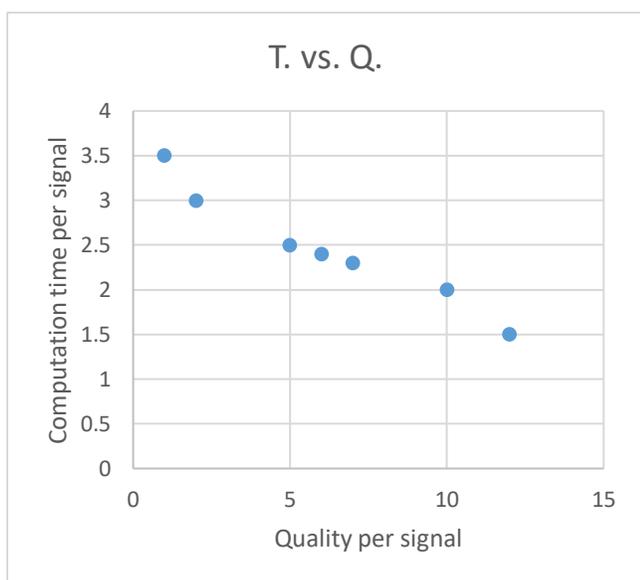

Figure 11: The relationship between the computation time and quality per signal

Then, three constraints were set for the maximum energy, time cost and the number of signals per type. The constraints were set so the total energy does not exceed ten thousand and the total computation time does not exceed two hundred and fifty microseconds. The programs were set so the number of signals per type doesn't exceed twenty. In other words, the program cannot select more than twenty signals from a certain type to find the maximum quality.

## VII. Testing methodology

The performance evaluation portion of this solution was done in three stages. The first stage is studying basic cases of parameters and constraints to find the maximum quality using the problem based, solver based and heuristic algorithm. In the first stage, only four different types of signals were created for testing. In every case the quality, time and energy specifications per signal were provided for each type. Then, the constraints were set. Finally, the objective of the algorithm is to find the optimal number of signals of each type that would result in the maximum total quality.

Once all three solutions are running and providing the correct solution which is the same for all of them. The second stage is to test and compare the running time for all three solutions. The programs were tested twenty times and the solution time for all the trials can be found in Table 6 below. All the results will be analyzed in the following section.

The third and final stage is to test the running time for all three proposed solution and compare them using more signal types. Three more types of signals were created for this problem. All the parameters for signals types 5, 6 and 7 are provided in Table 5. The programs were tested to find the maximum quality utilizing seven types of signals. Once they were running correctly, they were tested for the running time. The programs were again tested twenty times and the solution time for all the trials can be found in Table 6 below. All the results will be analyzed in the following section.

## VIII. Results

Integer linear programming was used in two approaches: problem-based and solver-based. The problem-based solution was introduced in the previous chapter. Solver-based solution was implemented and tested for the case with four signal types. The test was done twenty times and the results show improvement in the performance. As shown in Table 6, the performance was enhanced from 0.64 to 0.238 seconds. Then, the heuristic algorithm was implemented and tested. The solution time was improved to 0.178 seconds as shown in Table 6. All of this was implemented on MATLAB.

To test the solutions on a larger problem, they were tested again using seven signal types. The running time for the problem-based, solver-based and the heuristic algorithm were respectively 0.667, 0.271 and 0.183 seconds. The running time for each solution was more than the result for the first case. This is because it takes more time to solve a larger computational problem. A comparison the results of twenty different trials for each solution utilizing four or seven signals types are represented in Table 6 and Figure 12. Finally, the average of all the solution time results is shown in Table 6 and Figure 13.

| Trial | L.P. 4 types Problem based | L.P. 4 types Solver based | H.A. 4 types | L.P. 7 types Problem based | L.P. 7 types Solver based | H.A. 7 types |
|---|---|---|---|---|---|---|
| 1 | 0.548 | 0.111 | 0.100 | 0.668 | 0.197 | 0.193 |
| 2 | 0.642 | 0.165 | 0.204 | 0.649 | 0.279 | 0.189 |
| 3 | 0.601 | 0.130 | 0.156 | 0.564 | 0.438 | 0.178 |
| 4 | 0.685 | 0.265 | 0.184 | 0.824 | 0.289 | 0.171 |
| 5 | 0.617 | 0.232 | 0.189 | 0.624 | 0.236 | 0.182 |
| 6 | 0.600 | 0.251 | 0.192 | 0.676 | 0.267 | 0.178 |
| 7 | 0.675 | 0.243 | 0.170 | 0.713 | 0.296 | 0.172 |
| 8 | 0.632 | 0.249 | 0.168 | 0.635 | 0.276 | 0.186 |
| 9 | 0.649 | 0.229 | 0.189 | 0.648 | 0.265 | 0.178 |
| 10 | 0.651 | 0.243 | 0.169 | 0.619 | 0.293 | 0.224 |
| 11 | 0.659 | 0.254 | 0.189 | 0.667 | 0.283 | 0.186 |
| 12 | 0.621 | 0.265 | 0.182 | 0.667 | 0.339 | 0.175 |
| 13 | 0.661 | 0.270 | 0.188 | 0.658 | 0.259 | 0.192 |
| 14 | 0.646 | 0.256 | 0.179 | 0.647 | 0.228 | 0.179 |
| 15 | 0.629 | 0.264 | 0.183 | 0.661 | 0.258 | 0.182 |
| 16 | 0.629 | 0.300 | 0.173 | 0.678 | 0.235 | 0.184 |
| 17 | 0.658 | 0.261 | 0.192 | 0.762 | 0.253 | 0.186 |
| 18 | 0.695 | 0.261 | 0.204 | 0.690 | 0.238 | 0.177 |
| 19 | 0.661 | 0.257 | 0.175 | 0.644 | 0.242 | 0.165 |
| 20 | 0.645 | 0.252 | 0.177 | 0.649 | 0.245 | 0.175 |
| Avg. | 0.640 | 0.238 | 0.178 | 0.667 | 0.271 | 0.183 |

Table 6: Comparison between I.L.P. solutions and heuristic algorithm solution time results

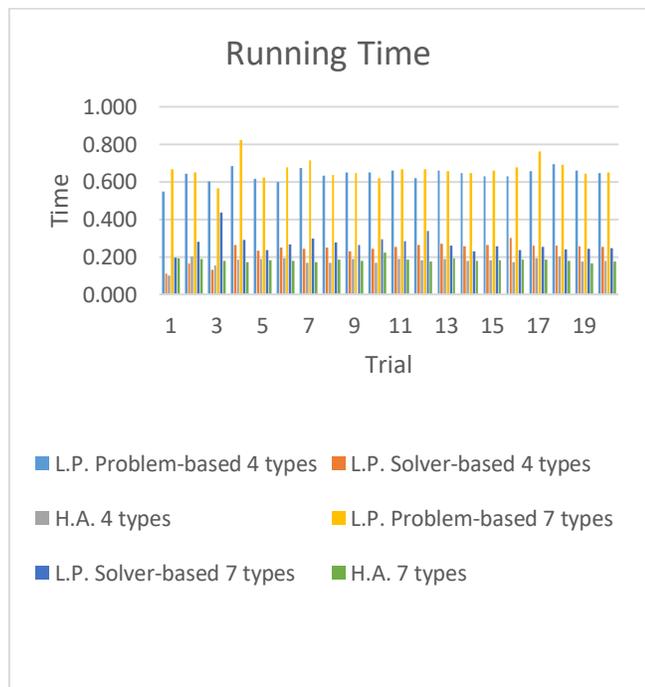

Figure 12: The solution time results

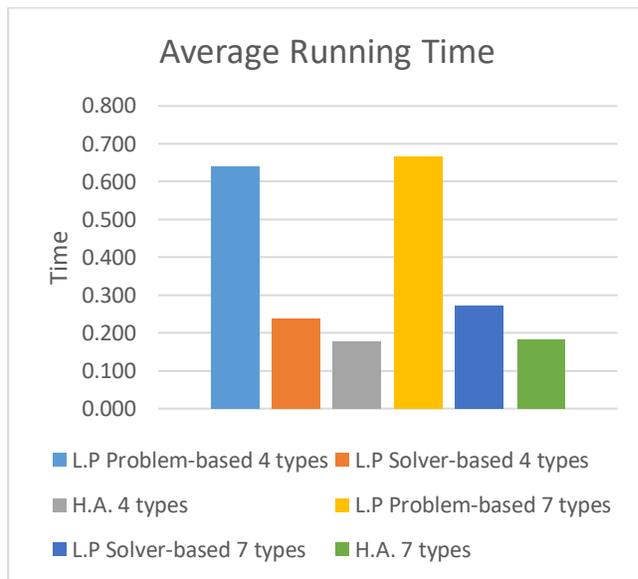

Figure 13: The average solution time results

## IX. Conclusion and Future Work

Integer linear programming was used in this paper to optimize sensing. Integer linear programming is a mathematical method to achieve the best outcome and it was used to select the best signals to be emitted and processed. Optimal sensing means acquiring more information with minimum energy and time costs which leads to more efficient processing of data.

The novelty of this work lies in the synthesis of an optimized signal mix using information such as but not limited to signal quality, energy and computing time. It was found that:

*A. The problem is easily formulated, and additional constraints can be simply added.*

*B. The problem can be solved by integer linear programming or heuristic algorithms.*

*C. The heuristic algorithm offers a significant reduction in computing time.*

*D. There are many applications in the sensing and telecommunications area for this concept. It can be applied to both active and passive sensing.*

Such intelligent generation and processing of signals is part of the well-recognized trend of increasing computer capabilities making possible useful applications.

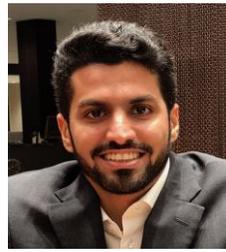

**Abdulaziz M. Alqarni** received a B.S.E.E. from Western Michigan University, Kalamazoo, MI, in 2014, M.S. degree in electrical engineering from Loyola Marymount University, Los Angeles, CA, and the Ph.D. from the Department of Electrical Engineering and Computer Engineering in Stony Brook University, Stony Brook, NY, in 2020.

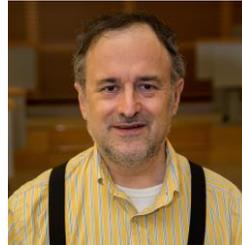

**Thomas G. Robertazzi** received the Ph.D. from Princeton University, Princeton, NJ, in 1981 and the B.E.E. from the Cooper Union, New York, NY in 1977.

He is presently a Professor in the Dept. of Electrical and Computer Engineering and an Affiliate Professor in the Dept. of Applied Mathematics and Statistics at Stony Brook University, Stony Brook N.Y. He has published in the areas of parallel processing, telecommunications, switching, queuing and Petri networks.

He has authored, co-authored or edited six books in the areas of performance evaluation, scheduling and network planning. He is a Fellow of the IEEE.